\documentclass[twocolumn,showpacs,prl]{revtex4}
\usepackage{graphicx}
\usepackage{graphicx}
\usepackage{epstopdf}
\usepackage{amsmath}
\usepackage{subfigure}
\usepackage{amssymb}
\usepackage{mdframed}
\usepackage{color}
\usepackage{lineno}

\makeatletter

\newcommand{\Rmnum}[1]{\expandafter\@slowromancap\romannumeral #1@}

\makeatother

\begin{document}

\title{{ Generic predictions of output probability based on complexities of inputs and outputs}}

\author{Kamaludin Dingle$^{1,2}$, Guillermo Valle P\'erez$^2$, Ard A. Louis$^2$}

\affiliation{$^1$Centre for Applied Mathematics and Bioinformatics,
Department of Mathematics and Natural Sciences,
Gulf University for Science and Technology, Kuwait,\\
$^2$Rudolf Peierls Centre for Theoretical Physics$\text{,}$ University of Oxford$\text{,}$ Parks Road,
Oxford, OX1 3PU, United Kingdom}

\date{\today}

\begin{abstract}\noindent
For a broad class of input-output maps,  arguments based on the coding theorem from algorithmic information theory (AIT) predict that  simple (low Kolmogorov complexity)  outputs are exponentially more likely to occur upon uniform random sampling of inputs than complex outputs are.
Here, we derive probability bounds that are based on the complexities of the inputs as well as the outputs, rather than just on the complexities of the outputs.  The more that outputs deviate from the coding theorem bound, the lower the complexity of their inputs.
Our new bounds  are tested for an RNA sequence to structure map, a finite state transducer and a perceptron.  These results open avenues for AIT to be more widely used in physics.
\end{abstract}


\maketitle
\twocolumngrid


Deep links between physics and theories of computation~\cite{mezard2009information,moore2011nature} are being increasingly exploited to uncover new fundamental physics and to provide novel insights into theories of computation.   For example, advances in understanding quantum entanglement are often expressed in sophisticated information theoretic language, while providing new results in computational complexity theory such as polynomial time algorithms for integer factorization~\cite{shor1994algorithms}.
These connections 
 are typically expressed in terms of Shannon information, with  its natural analogy with thermodynamic entropy.

 There is, however, another branch of information theory, called algorithmic information theory (AIT)~\cite{li2008introduction}, which is concerned with the information content of individual objects. It has been much less applied in physics (although notable exceptions occur, see~\cite{devine2014algorithmic} for a recent overview).  Reasons for this relative lack of attention include that AIT's central concept, the Kolmogorov complexity $K_U(x)$ of a string $x$, defined as the length of the shortest program that generates  $x$ on a universal Turing machine (UTM) $U$, is formally uncomputable due to its link to the famous halting problem of UTMs~\cite{turing1936computable}.  Moreover,  many important results, such as the invariance theorem which states that for two UTMs $U$ and $W$, the Kolmogorov complexities $K_U(x) = K_W(x) + \mathcal{O}(1)$ are equivalent, hold asymptotically up to  $\mathcal{O}(1)$ terms that are independent of $x$, but not always well understood, and therefore hard to control.

Another reason applications of AIT to many practical problems have been hindered can be understood in terms of hierarchies of computing power.  For example, one of the oldest such categorisations, the Chomsky hierarchy~\cite{chomsky1956three}, ranks automata into four different classes, of which the UTMs are the most powerful, and finite state machines (FSMs) are the least.  Many key results in AIT are derived by exploiting  the power of UTMs.   Interestingly, if physical processes can be mapped onto UTMs, then certain properties can be shown to be uncomputable~\cite{lloyd1993quantum,cubitt2015undecidability}. However, many problems in physics are fully computable, and therefore lower  on the Chomsky hierarchy than UTMs. For example, finite Markov processes are equivalent to FSMs, and  RNA secondary structure (SS) folding algorithms can be recast as context-free grammars, the second level in the hiearchy.  Thus, an important cluster of questions for applications of AIT revolve around extending its methods to processes lower in computational power than UTMs.

To explore ways of moving beyond these limitations and towards practical applications, we consider here one of the most iconic results of AIT, namely the coding theorem of Solomonoff and Levin~\cite{solomonoff1964formal,levin1974laws}, which predicts that upon randomly chosen programs, the probability $P_U(x)$ that a universal Turing machine (UTM) 
generates output $x$ can be bounded as $ 2^{-K(x)} \leq P(x) \leq 2^{-K(x)+{\mathcal O}(1)}$. 
 Given this profound prediction of a general exponential bias towards simplicity  (low Kolmogorov complexity) one might have expected widespread study and applications in science and engineering.  This has not been the case because  the theorem unfortunately suffers from the general issues of AIT described above (see however~\cite{delahaye2012numerical,zenil2014correlation,soler2014calculating} for important attempts to apply the full coding theorem). 

 Nevertheless,  it has recently been shown~\cite{dingle2018input,zenil2019coding} that a related exponential bias towards low complexity outputs obtains for a range of non-universal input-output maps $f:I \rightarrow O$ that are lower on the Chomsky hierarchy than UTMs.
 \begin{figure*}[htp]
 \subfigure[]{\label{fig:edge-d}\includegraphics[height=4cm,width=5cm]{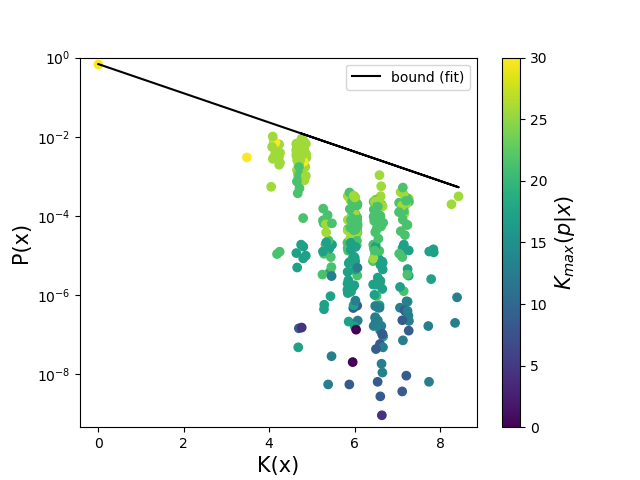}}
\subfigure[]{\label{fig:simpbias_FST}\includegraphics[height=4cm,width=5cm]{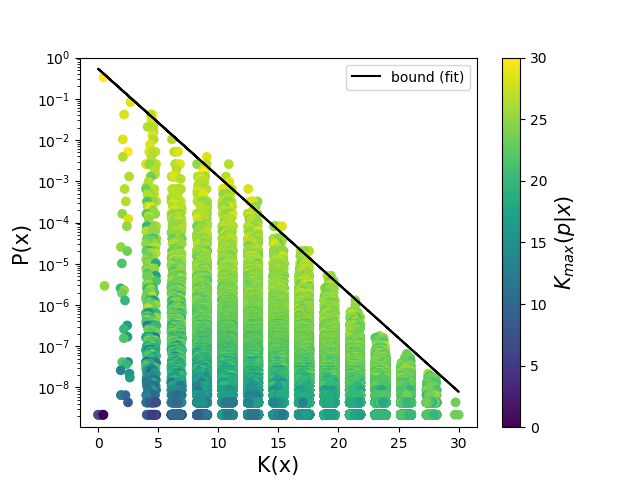}}
\subfigure[]{\label{fig:simpbias_perceptron}\includegraphics[height=4cm,width=5cm]{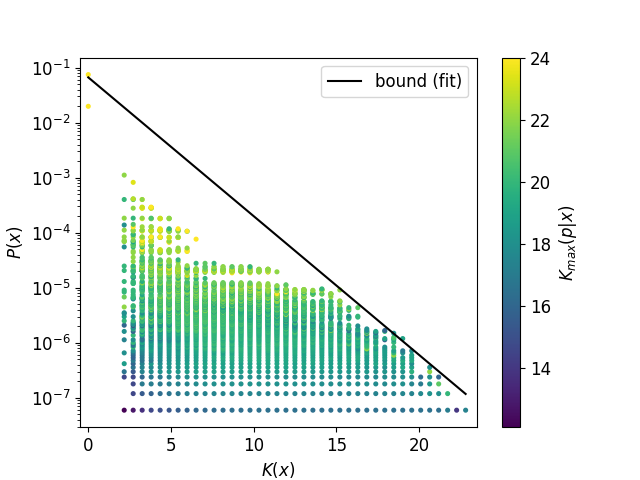}}
\subfigure[]{\label{fig:edge-d}\includegraphics[height=4cm,width=5cm]{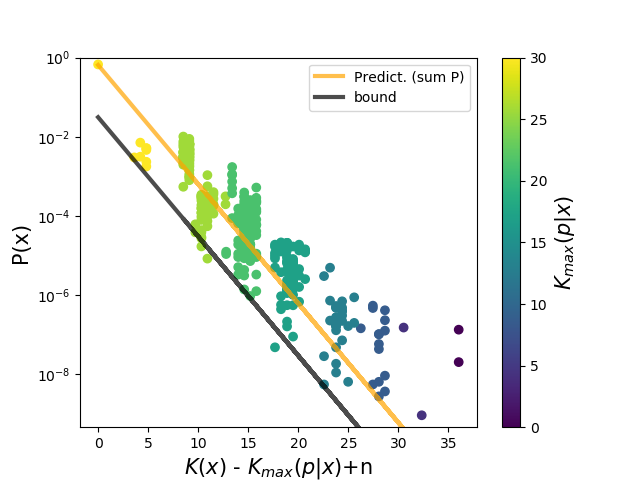}}
\subfigure[]{\label{fig:edge-d}\includegraphics[height=4cm,width=5cm]{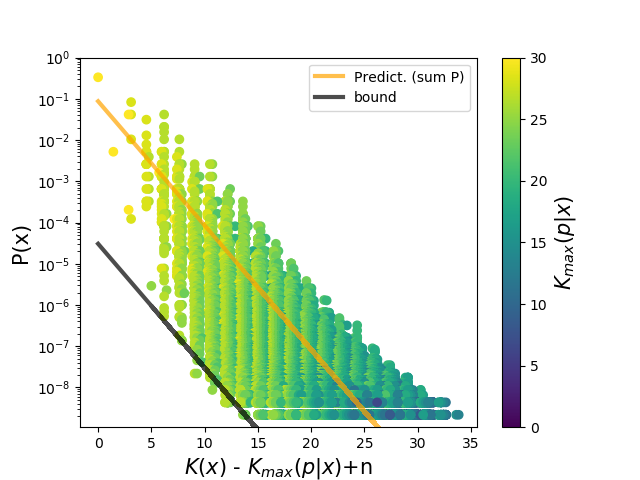}}
\subfigure[]{\label{fig:edge-d}\includegraphics[height=4cm,width=5cm]{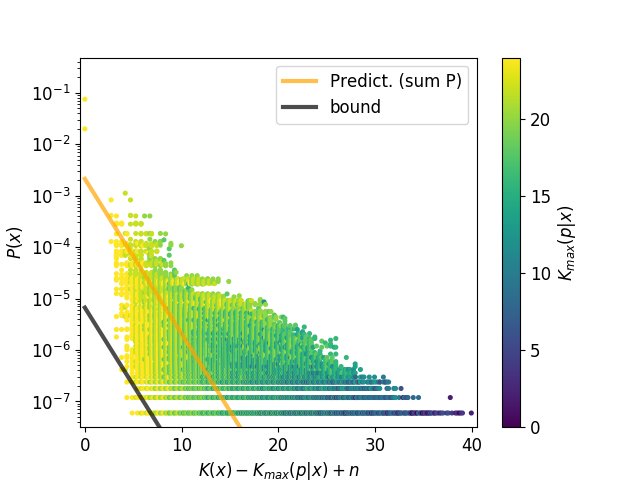}}

\caption{The probability $P(x)$ that a particular output arises upon random sampling of inputs  versus output complexity $\tilde{K}(x)$ shows clear simplicity bias for: (a) A length $L=15$ RNA sequence to SS mapping, (b) An FST, sampled over all $2^{30}$ binary inputs of length $30$, and (c) A 7-input perceptron with weights discretised to 3 bits.    The black solid line is the simplicity bias bound~(\ref{eq:SB}) (with $a$ and $b$ fit). For all these maps high complexity outputs occur with low probability.
  The outputs are colour coded by the maximum complexity $K_{\text{\rm max}}(p|n)$  of the set of inputs mapping to output $x$.  Outputs further from the bound have lower input complexities.
Figs (d)    length $L=15$ RNA, (e) the FST and (d) the perceptron, show the data plotted for the lower bound~(\ref{eq:bound3_k1}) (black line)  with only the intercept fit to the data, the slope is a prediction.   The orange line is using  Eq~(\ref{eq:bound3_k1}) with a normalised probability for a  parameter free predictor.  Including the complexity of the input through $K_{\text{\rm max}}(p|n)$ reduces the spread in the data, and so provides more predictive power than $K(x)$ alone.
}\label{fig:1}
\end{figure*}
In particular, an upper bound on the probability $P(x)$ that an output obtains upon uniform random sampling of inputs,
\begin{equation}
P(x)\leq 2^{-a\tilde K(x)-b}\label{eq:SB}\,
\end{equation}
was recently derived~\cite{dingle2018input} using a computable approximation $\tilde K(x)$ to the Kolmogorov complexity of $x$, typically calculated using lossless compression techniques.   Here  $a$ and $b$ are constants that are independent of $x$ and which can often be determined from some basic  information about the map.  The so-called {\em simplicity bias} bound~(\ref{eq:SB}) holds for computable maps $f$ where the number of inputs $N_I$ is much greater than the number of outputs $N_O$ and the map $K(f)$  is simple, meaning  that asymptotically $K(f) + K(n) \ll K(x) + \mathcal{O}(1)$ for a typical output $x$, where $n$ specifies the size of $N_I$, e.g.\ $N_I = 2^n$.  Eq.~(\ref{eq:SB})  typically works better for larger $N_I$ and $N_O$.  Approximating the true Kolmogorov complexity also means that the bound shouldn't work for maps where a significant fraction of outputs have complexities that are not qualitatively captured by compression based approximations.   For example many pseudo random-number generators are designed to produce outputs that appear to be complex when measured by compression or other types of Kolmogorov  complexity approximations. Yet these outputs must have low  $K(x)$ because they are generated by relatively simple algorithms with short descriptions.
Nevertheless, it has been shown that the bound~(\ref{eq:SB}) works remarkably well for a wide class of input-output maps,   ranging from  the sequence to RNA secondary structure map, to systems of coupled ordinary differential equations, to a stochastic financial trading model, to the parameter-function map for several classes of deep neural networks~\cite{dingle2018input,perez2018deep,mingard2019neural}.

The simplicity bias bound~(\ref{eq:SB}) predicts that high $P(x)$ outputs will be simple, and that complex outputs will have a low $P(x)$.   But, in sharp contrast to the full AIT coding theorem, it doesn't have a lower bound, allowing low $\tilde{K}(x)$ outputs with low $P(x)$ that are far from the bound.   Indeed, this behaviour is generically observed for many (non-universal) maps~\cite{dingle2018input,perez2018deep} (see also Fig 1), but should not be the case for UTMs that  obey the full coding theorem.
Understanding the behaviour of outputs far from the bound should shed light on fundamental differences between UTMs and maps with less computational power that are lower on the Chomsky hierarchy, and may  open up avenues for wider applications of AIT in physics.

With this challenge in mind, we take an approach that contrasts with the traditional coding theorem of AIT or with the simplicity bias bound, which only consider the complexity of the outputs.  Instead, we derive bounds that also take into account the complexity of the inputs that generate a particular output $x$.   While this approach is not possible for UTMs, since the halting problem means one cannot enumerate all inputs~\cite{li2008introduction}, and so averages over input complexity cannot be calculated,  it can be achieved for non-UTM maps.  Among our main results, we show that the further outputs are from the simplicity bias bound~(\ref{eq:SB}), the lower the complexity of the set of inputs. Since, by simple counting arguments, most strings are complex~\cite{li2008introduction}, the cumulative probability of outputs far from the bound is therefore limited. We also show that by combining the complexities of the output with that of the inputs, we can obtain better bounds on and estimates of $P(x)$.

Whether such bounds nevertheless have real predictive power needs to be tested empirically.
Because input based bounds typically need exhaustive sampling, full testing is only possible for smaller systems,  which restricts us here  to maps where finite size effects may still play a role~\cite{dingle2018input}.
We test our bounds on three systems,
 the famous RNA sequence to secondary structure map (which falls into the context-free class in the Chomsky hierarchy), here for a relatively small size with length $L=15$ sequences, a finite state transducer (FST), a very simple input-output map that is lowest on the Chomsky hierarchy~\cite{chomsky1956three}, with length $L=30$ binary inputs, and finally the parameter-function map~\cite{valle2018deep,mingard2019neural} of a perceptron~\cite{rosenblatt1958perceptron} with discretized weights to allow complexities of inputs to be calculated. The preceptron plays a key role in deep learning neural network architectures ~\cite{lecun2015deep}.  Nevertheless, as can be seen in Fig.~\ref{fig:1}(a-c) all three maps exhibit simplicity bias predicted by Eq~(\ref{eq:SB}),  even if they are relatively small. In Ref.~\cite{dingle2018input}, much cleaner simplicity bias behaviour can be observed  for larger RNA maps, but these are too big to exhaustively sample inputs.  Similarly, cleaner simplicity bias behaviour occurs for the undiscretised perceptron~\cite{mingard2019neural}, but then it is hard to analyse the complexity of the inputs.

 Fig.~\ref{fig:1}(a-c) shows that the complexity of the input strings that generate each output $x$ decreases for further distances from the simplicity bias bound.  This is the kind of phenomenon that the we will attempt to explain.

To study input based bounds, consider a map $f:I\rightarrow O$ between $N_I$ inputs and $N_O$ outputs that satisfies the requirements for simplicity bias~\cite{dingle2018input}.  Let $f(p)=x$, where $p$ is some input program $p \in I$  producing output $x \in O$. For simplicity let $p\in \{0,1\}^n$, so that all inputs have length $n$ and $N_I = 2^n$ (this restriction can be relaxed later). Define  $f^{-1}(x)$
to be
the set of all the inputs that map to $x$, so that the probability that $x$ obtains upon sampling inputs uniformly at random is
\begin{equation}\label{eq:A}
P(x)=\frac{|f^{-1}(x)|}{2^n}\,
\end{equation}
Any arbitrary input $p$ can be described using the following $ \mathcal{O}(1)$ procedure~\cite{dingle2018input}:  Assuming $f$ and $n$ are given,  first enumerate all $2^n$ inputs and map them to outputs using $f$.
The index of a specific input $p$ within the set $f^{-1}(x)$ can be described using at most $\log_2(|f^{-1}(x)|)$ bits. 
   In other words, this procedure identifies each input by first finding the output $x$ it maps to, and then finding its label within the set  $f^{-1}(x)$.  Given $f$ and $n$, an output $x=f(p)$ can be described using $K(x|f,n) + \mathcal{O}(1)$ bits~\cite{dingle2018input}. Thus, the Kolmogorov complexity of $p$, given $f$ and $n$ can be bounded as:
\begin{equation}
K(p|f,n) \leq K(x|f,n)+\log_2(|f^{-1}(x)|) +\mathcal{O}(1) \label{eq:bound2_p}.
\end{equation}

We note that this bound holds in principle for all $p$, but that it is tightest for $K_{\rm max}(p|x) \equiv \max_p\{K(p|f,n )\}$ for $p \in f^{-1}(x)$.
More generally, we can  expect these bounds to be fairly tight for the maximum complexity $K_{\text{\rm max}}(p|f,n)$ of inputs due to the following argument.  First note that
\begin{equation} K_{\text{\rm max}}(p|f,n)\geq \log_2(|f^{-1}(x)|) + {\mathcal O}(1)
\end{equation}  because any set of $|f^{-1}(x)|$ different elements must have strings of at least this complexity. Next,
 \begin{equation} \label{eq:px}
 K(x|f,n)\leq K(p|f,n)+{\mathcal O}(1)
 \end{equation}  because each $p$ can be used to generate $x$. Therefore:
\begin{equation}
\max(K(x|f,n),\log_2(|f^{-1}(x)|))\leq K_{\text{\rm max}}(p|f,n)+{\mathcal O}(1),
\end{equation}
so the bound~(\ref{eq:bound2_p})  cannot be too weak. In the worst case scenario, where $K_{\text{\rm max}}(p|n) \approx \log_2(|f^{-1}(x)|) \approx K(x|f,n)$,  the right hand side of the bound~(\ref{eq:bound2_p}) is approximately twice the left hand side (up to additive ${\mathcal O}(1)$ terms).  It is tighter  if either $K(x|f,n)$ is small, or if $K(x|f,n)$ is big relative to  $\log_2(|f^{-1}(x)|)$.  As is often the case for AIT predictions,  the stronger the constraint/prediction, the more likely it is to be observed in practice, because, for example, the ${\mathcal O}(1)$ terms are less likely to drown out the effects.

\begin{figure*}[htp]
\begin{center}
\subfigure[]{\label{fig:edge-d}\includegraphics[height=4cm,width=5cm]{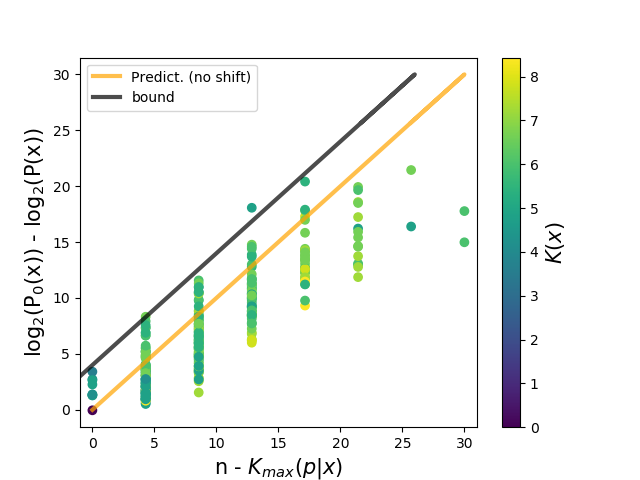}}
\subfigure[]{\label{fig:edge-d}\includegraphics[height=4cm,width=5cm]{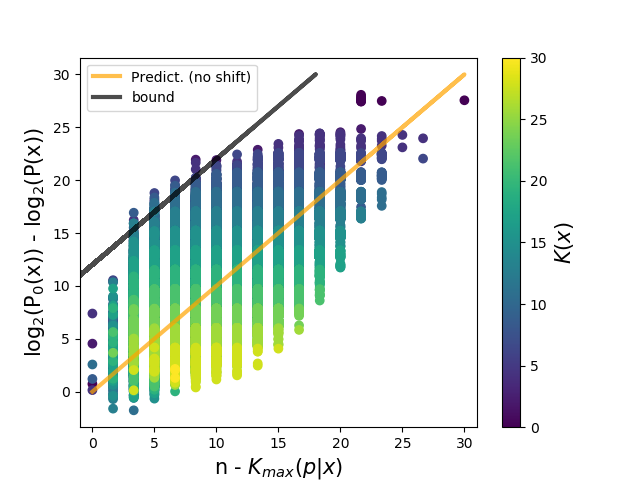}}
\subfigure[]{\label{fig:edge-d}\includegraphics[height=4cm,width=5cm]{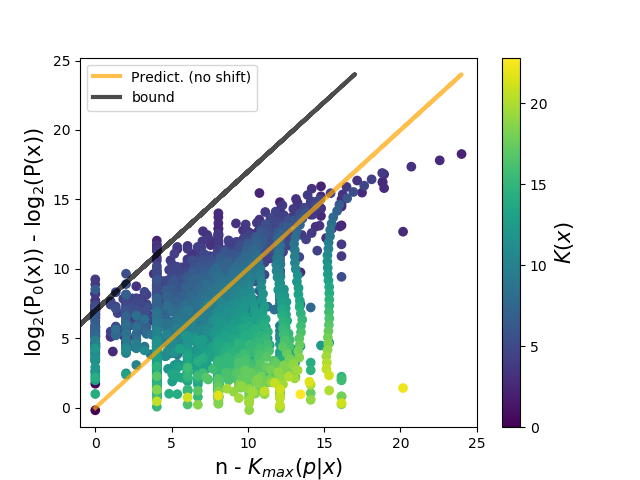}}
\end{center}
\caption{Deviation of $P(x)$ from the simplicity bias upper bound~(\ref{eq:SB})) increases with increasing randomness deficit   $\delta_{\rm max}(x) =n - K_{\text{\rm max}}(p|n)$ for (a) $L=15$ RNA, (b) $L=30$ FST, (c)  perceptron with weights discretised to 4 bits. For the perceptron, all functions with the same $P(x)$ and $K(x)$ are averaged together to reduce scatter.   Points are colour coded by output complexity $K(x)$.
For the upper bound~(\ref{eq:bound3_p}) (black line) we fit the intercept, but the slope is a prediction, if we treat it as a normalised probability we obtain the orange line which is a direct prediction with no free parameters.   
}
\label{fig:newbounds}
\end{figure*}

By combining with Eq.~(\ref{eq:A}), the bound~(\ref{eq:bound2_p}) can  be rewritten in two complementary ways.
Firstly, a lower bound on $P(x)$ can be derived of the form:
\begin{equation}\label{eq:bound3_k}
P(x)  \geq 2^{ - K(x|f,n) -  [n - K(p|f,n)] +\mathcal{O}(1)}
\end{equation}
$\forall p\in f^{-1}(x)$
which complements the simplicity bias upper bound~(\ref{eq:SB}). This bound is tightest for $K_{\text{\rm max}}(p|n)$.

In Ref.\ ~\cite{dingle2018input} it was shown that $P(x) \leq 2^{-K(x|f,n) +\mathcal{O}(1)}$ by using a similar counting argument to that used above, together with a Shannon-Fano-Elias code procedure. Similar results can be found in standard works~\cite{li2008introduction,gacs1988lecture}. A key step is to move from the conditional complexity to one that is independent of the map and of $n$.  If $f$ is simple, then the explicit dependence on $n$ and $f$ can be removed by noting that since $K(x) \leq K(x|f,n) + K(f) + K(n) + \mathcal{O}(1)$, and $K(x|f,n) \leq K(x)  + \mathcal{O}(1)$ then $K(x|f,n) \approx K(x) +  \mathcal{O}(1)$. In  Eq.~(\ref{eq:SB}) this is further approximated as $K(x|f,n) +\mathcal{O}(1) \approx a \tilde{K}(x) + b$, leading to a practically useable upper bound.  The same argument can be used to remove explicit dependence on $n$ and $f$ for $K(p|f,n)$.

If we define a maximum randomness deficit  $\delta_{\rm max}(x) =n - K_{\text{\rm max}}(p|n)$, then this tightest version of bound~(\ref{eq:bound3_k})  can be written in a simpler form as
\begin{equation} \label{eq:bound3_k1}
P(x)  \geq 2^{ - a\tilde{K}(x)  + b - \delta_{\rm max}(x)  +\mathcal{O}(1)}
\end{equation}
In Figs.~1~(d-f) we plot this lower bound for all three maps studied. Throughout the paper, we use a scaled complexity measure, which ensures that $\tilde{K}(x)$ ranges between $\approx$0 and $\approx$$n$ bits, for strings of length $n$, as expected for Kolmogorov complexity. See Methods for more details.

     When comparing the data in Figs.~1~(d-f) to Figs.~1~(a-c), it is clear that including the input complexities reduces the spread in the data for RNA and the FST, although for the perceptron model the difference is less pronounced.  This success suggests using the bound~(\ref{eq:bound3_k1}) as a predictor $P(x) \approx  2^{ - K(x|f,n) - \delta_{\rm max}(x) }$, with the additional constraint   that $\sum_x P(x) = 1$ to normalise it.   As can be seen in Figs.~1~(d-f), this simple procedure works reasonably well, showing that the input complexity provides additional predictive power to estimate $P(x)$ from some very generic properties of the inputs and outputs.

A second, complimentary way that bound~(\ref{eq:bound2_p})  can be expressed is in terms of how far $P(x)$ differs from the simplicity bias bound~(\ref{eq:SB}):
\begin{equation}\label{eq:bound3_p}
\left[\log_2(P_0(x)) - \log_2(P(x))\right]  \leq  [n - K(p|f,n)] +\mathcal{O}(1)
\end{equation}
where $P_0(x) = 2^{-K(x|f,n)} \approx 2^{- a \tilde{K}(x) + b}$ is the upper bound~(\ref{eq:SB}) shown in Figs~1~(a-c).

 For a random input $p$, with high probability we expect $K(p|f,n) = n + \mathcal{O}(1)$~\cite{li2008introduction}. Thus,  eqs.~(\ref{eq:bound3_k}) and~(\ref{eq:bound3_p}) immediately imply that large deviations from the simplicity bias bound~(\ref{eq:SB}) are only possible with highly non-random inputs with a large randomness deficit $\delta_{\rm max}(x)$.

 In Fig.~\ref{fig:newbounds}(a)-(c) we directly examine bound~(\ref{eq:bound3_p}),  showing explicitly the prediction that a drop of probability $P(x)$ by $\Delta$ bits from the simplicity bias bound (~\ref{eq:SB}) corresponds to a $\Delta$ bit randomness deficit in the set of inputs.

   Simple counting arguments can be used to show that the number of non-random inputs is a small fraction of the total number of inputs~\cite{chaitin1974information}. For example, for binary strings of length $n$, with $N_I = 2^n$, the number of inputs with complexity $K=n- \delta$ is approximately  $2^{-\delta}N_I$.
 If we define a set  $\mathcal{D}(f)$ of all outputs $x_i$  that satisfy   $(\log_2(P_0(x_i)) - \log_2(P_(x_i))) \geq \Delta$, i.e.\ the set of all outputs for which $\log_2 P(x)$ is at least $\Delta$ bits below the simplicity bias bound~(\ref{eq:SB}), then this counting argument leads to the following cumulative bound:
 \begin{equation}
 \sum_{x \in \mathcal{D}(f)}   P(x) \leq 2^{-\Delta +1 + \mathcal{O}(1)}\label{eq:sumP_delta}
 \end{equation}
 which predicts that, upon randomly sampling inputs, most of the probability weight is for outputs with $P(x)$ relatively close to the upper bound.   There may be many outputs that are far from the bound, but their cumulative probability drops off exponentially the further they are from the bound because the number of simple inputs is exponentially limited.   Note that this argument is for a cumulative probability over all inputs.  It does not predict that for a given complexity $K(x)$, that the outputs should all be near the bound.   In that sense this lower bound is not like that of the original coding theorem which holds for any output $x$.

\begin{figure*}[htp]
\begin{center}
\subfigure[]{\label{fig:edge-d}\includegraphics[height=4cm,width=4.4cm]{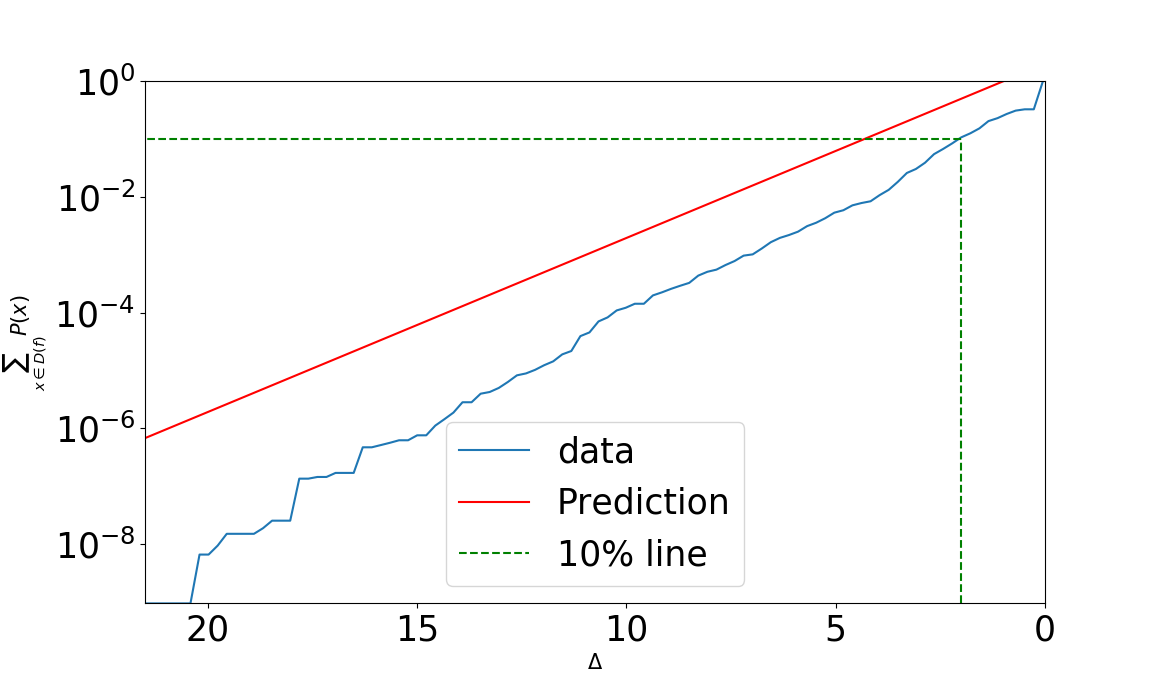}}
\subfigure[]{\label{fig:edge-d}\includegraphics[height=4cm,width=4.4cm]{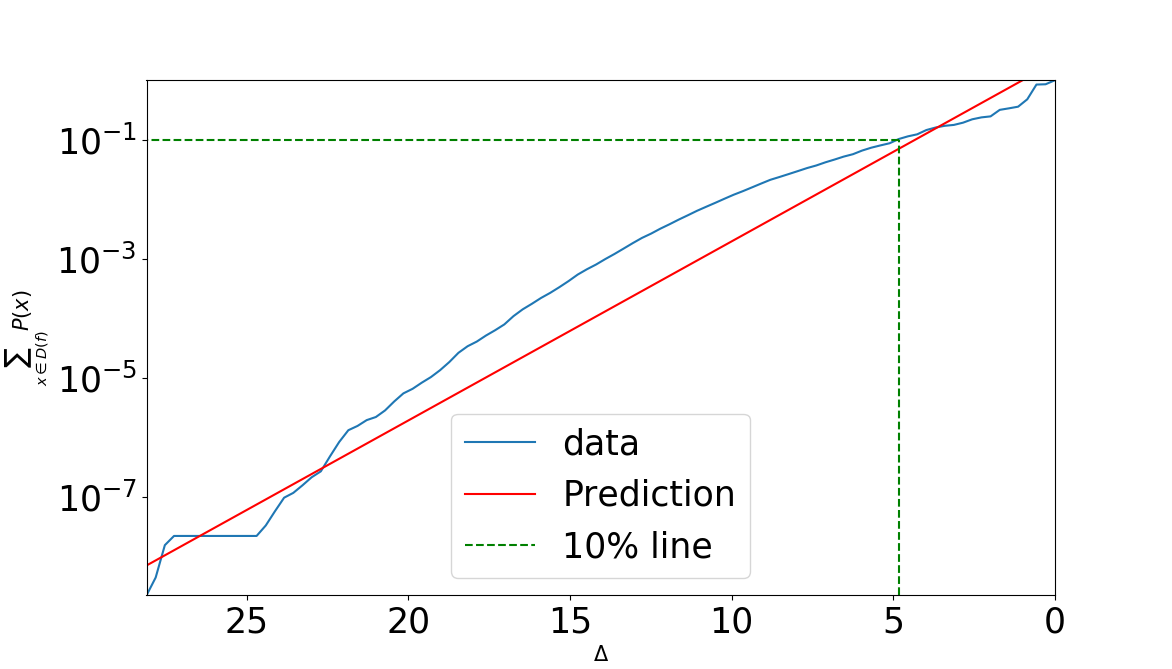}}
\subfigure[]{\label{fig:edge-d}\includegraphics[height=4cm,width=4.4cm]{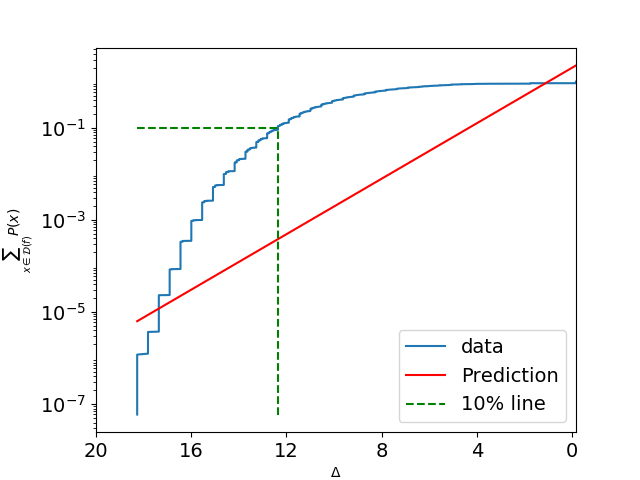}}
\subfigure[]{\label{fig:edge-d}\includegraphics[height=4cm,width=4.4cm]{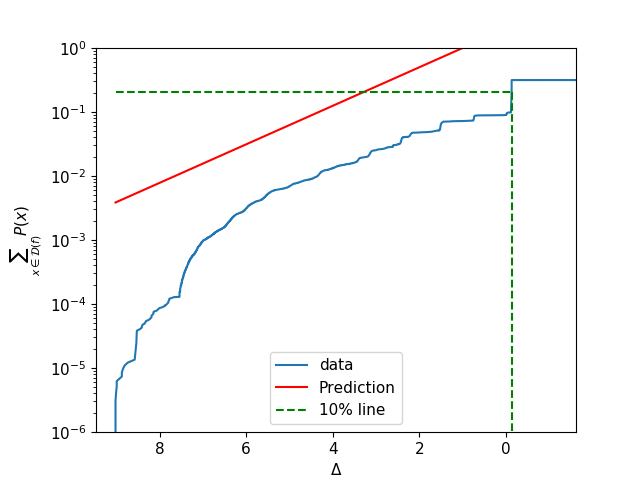}}
\subfigure[]{\label{fig:edge-d}\includegraphics[height=4cm,width=4.4cm]{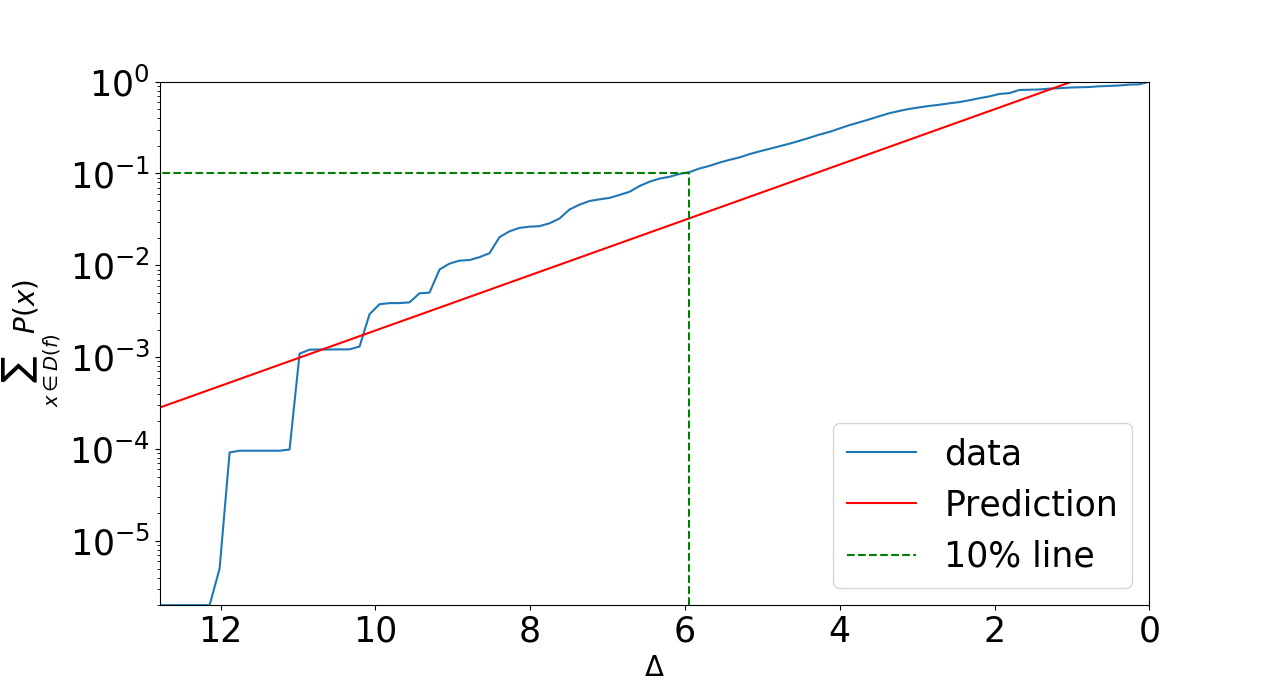}}
\subfigure[]{\label{fig:edge-d}\includegraphics[height=4cm,width=4.4cm]{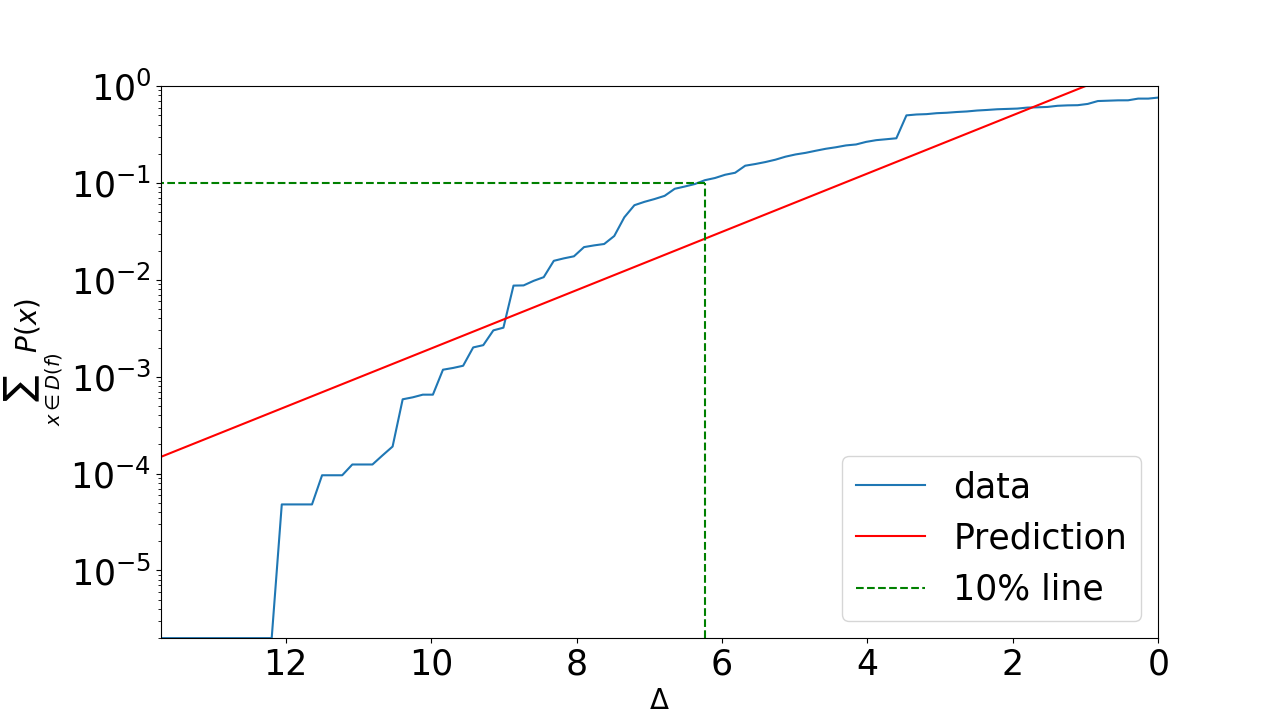}}
\subfigure[]{\label{fig:edge-d}\includegraphics[height=4cm,width=4.4cm]{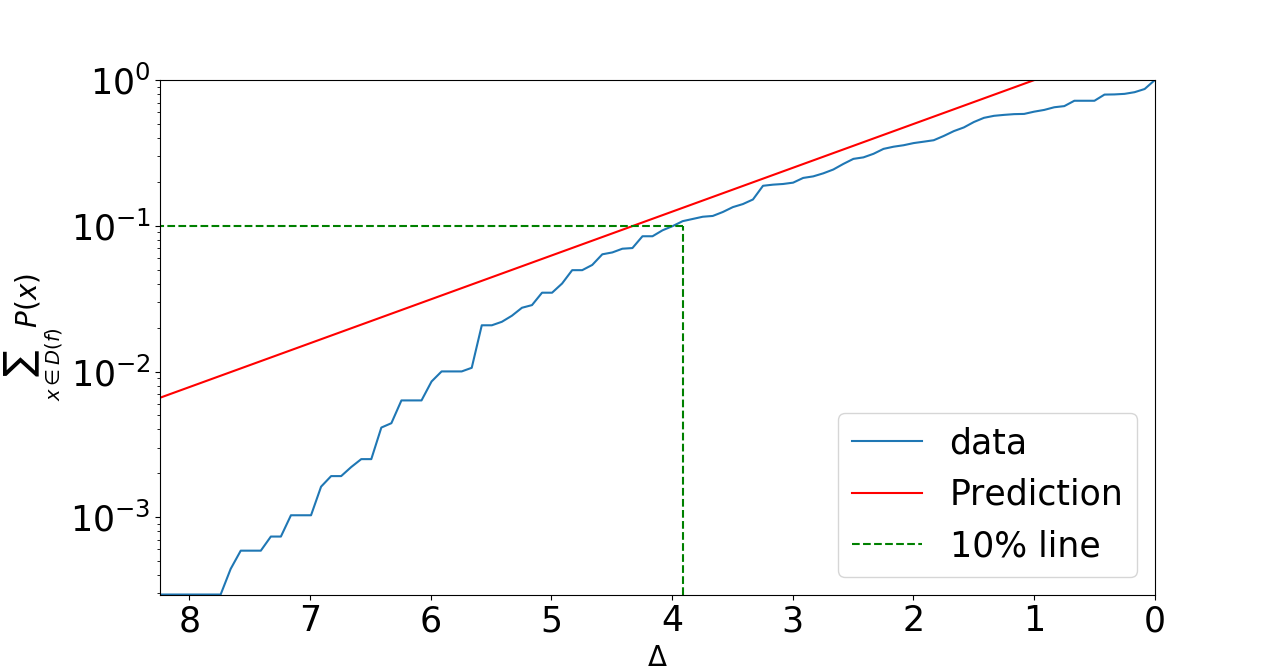}}
\subfigure[]{\label{fig:edge-d}\includegraphics[height=4cm,width=4.4cm]{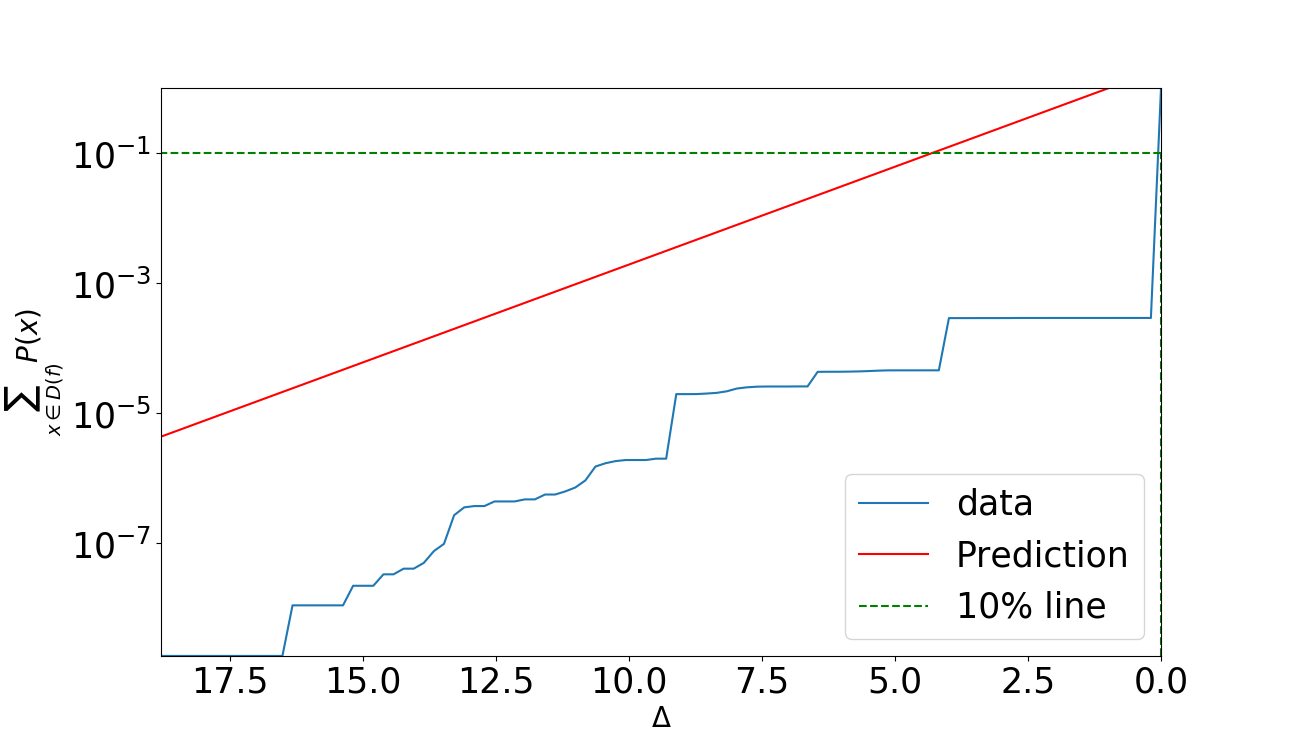}}
\end{center}
\caption{The cumulative probability versus the distance from the bound $\Delta$ correlates with the 
the cumulative bound~(\ref{eq:sumP_delta}) (red line) for (a) L=15 RNA and (b) L=30 FST  (c) Perceptron. (d) fully connected 2 layer neural network~from~\cite{valle2018deep}, (e) coarse-grained ordinary differential equation map from~\cite{dingle2018input}, (f) Ornstein-Uhlenbeck financial model from~\cite{dingle2018input}, (g) L-systems from~\cite{dingle2018input}, (h) simple matrix map from~\cite{dingle2018input}.  The solid red line is the prediction  $2^{-\Delta+1}$ from Eq.~(\ref{eq:sumP_delta}), the dashed line denotes $10\%$ cumulative probability.}
\label{fig:sumP_delta},
\end{figure*}

Bound~(\ref{eq:sumP_delta}) does not need an exhaustive enumeration to be tested.   In Fig.~\ref{fig:sumP_delta} we show this bound for a series of different maps, including many maps from~\cite{dingle2018input}. The cumulative probability weight scales roughly as expected, implying that most of the probability weight is relatively close to the bound (at least on a log scale).

What is the physical  nature of these low complexity, low probability outputs that occur far from the bound?  They must arise in one way or another from the lower computational power of these maps, since they don't occur in the full AIT coding theory. Low complexity, low probability outputs correspond to output patterns which are simple, but which the given computable map is not good at generating.

In RNA it is easy to construct outputs which are simple but will have low probability. 
Compare two $L=15$ structures  $S_1=$((.(.(...).).)). and $S_2=$.((.((...)).))., which are both symmetric and thus have a relatively low complexity $K(S_1)=K(S_2)$ $=21.4$. Nevertheless they have a significant difference in probability, $P(S_2)/P(S_1) \approx 560$ because $S_1$ has several single bonds, which is much harder to make according to the biophysics of RNA.  Only specially ordered input sequences can make $S_1$, in other words  they are simple, with $K_{\text{\rm max}}(p|n) =8.6$.   By contrast, the inputs of $S_2$ are much higher at $K_{\text{\rm max}}(p|n) = 21.4$ because they need to be constrained less to produce this structure.
This example illustrates how the system specific details of the RNA map can unfavourably bias away from some outputs due to a system specific constraint.

Similar examples of system specific constraint for the FST and perceptron can be found in the SI. We hypothesise that such low complexity, low probability structures highlight specific non-universal aspects of the maps, and extra information (in the form of  a reduced set of inputs) are needed to generate such structures.

In conclusion, it is striking that bounds based simply on the complexity of the inputs and outputs can make powerful and general predictions for such a wide range of systems.
 Although the arguments used to derive them suffer from the well known problems  -- e.g.\ the presence of uncomputable Kolmogorov complexities and unknown  ${\mathcal O}(1)$ terms  -- that have led to the general neglect of AIT in the physics literature,  the bounds are undoubtably successful.  It appears that, just as is found in other areas of physics, these relationships hold  well outside of the asymptotic regime where they can be prove to be correct.  This practical success opens up the promise of using such AIT based techniques to derive other results for computable maps from across physics.

Many new questions arise. Can it be proven when  the $\mathcal{O}(1)$ terms are relatively unimportant?  Why do our rather simple approximations to $K(x)$ work?   It would be interesting to find maps where these classical objections to the practical use of AIT are important.  There may also be  connections between our work and finite state complexity~\cite{calude2011finite} or minimum description length~\cite{grunwald2019minimum} approaches.
  Progress in these domains should  generate new fundamental understandings of the physics of information.

\begin{acknowledgments} K.D. acknowledges partial financial support from the Kuwait Foundation for the Advancement of Sciences (KFAS) grant number P115-12SL-06. G.V.P. acknowledges financial support from EPSRC through grant EP/G03706X/1.
\end{acknowledgments}


\begin{thebibliography}{25}
\expandafter\ifx\csname natexlab\endcsname\relax\def\natexlab#1{#1}\fi
\expandafter\ifx\csname bibnamefont\endcsname\relax
  \def\bibnamefont#1{#1}\fi
\expandafter\ifx\csname bibfnamefont\endcsname\relax
  \def\bibfnamefont#1{#1}\fi
\expandafter\ifx\csname citenamefont\endcsname\relax
  \def\citenamefont#1{#1}\fi
\expandafter\ifx\csname url\endcsname\relax
  \def\url#1{\texttt{#1}}\fi
\expandafter\ifx\csname urlprefix\endcsname\relax\def\urlprefix{URL }\fi
\providecommand{\bibinfo}[2]{#2}
\providecommand{\eprint}[2][]{\url{#2}}

\bibitem[{\citenamefont{Mezard and Montanari}(2009)}]{mezard2009information}
\bibinfo{author}{\bibfnamefont{M.}~\bibnamefont{Mezard}} \bibnamefont{and}
  \bibinfo{author}{\bibfnamefont{A.}~\bibnamefont{Montanari}},
  \emph{\bibinfo{title}{Information, physics, and computation}}
  (\bibinfo{publisher}{Oxford University Press, USA}, \bibinfo{year}{2009}).

\bibitem[{\citenamefont{Moore and Mertens}(2011)}]{moore2011nature}
\bibinfo{author}{\bibfnamefont{C.}~\bibnamefont{Moore}} \bibnamefont{and}
  \bibinfo{author}{\bibfnamefont{S.}~\bibnamefont{Mertens}},
  \emph{\bibinfo{title}{The nature of computation}} (\bibinfo{publisher}{OUP
  Oxford}, \bibinfo{year}{2011}).

\bibitem[{\citenamefont{Shor}(1994)}]{shor1994algorithms}
\bibinfo{author}{\bibfnamefont{P.~W.} \bibnamefont{Shor}}, in
  \emph{\bibinfo{booktitle}{Proceedings 35th annual symposium on foundations of
  computer science}} (\bibinfo{organization}{Ieee}, \bibinfo{year}{1994}), pp.
  \bibinfo{pages}{124--134}.

\bibitem[{\citenamefont{Li and Vitanyi}(2008)}]{li2008introduction}
\bibinfo{author}{\bibfnamefont{M.}~\bibnamefont{Li}} \bibnamefont{and}
  \bibinfo{author}{\bibfnamefont{P.}~\bibnamefont{Vitanyi}},
  \emph{\bibinfo{title}{An introduction to Kolmogorov complexity and its
  applications}} (\bibinfo{publisher}{Springer-Verlag New York Inc},
  \bibinfo{year}{2008}).

\bibitem[{\citenamefont{Devine}(2014)}]{devine2014algorithmic}
\bibinfo{author}{\bibfnamefont{S.}~\bibnamefont{Devine}},
  \emph{\bibinfo{title}{Algorithmic information theory: Review for physicists
  and natural scientists}} (\bibinfo{year}{2014}).

\bibitem[{\citenamefont{Turing}(1936)}]{turing1936computable}
\bibinfo{author}{\bibfnamefont{A.~M.} \bibnamefont{Turing}},
  \bibinfo{journal}{J. of Math} \textbf{\bibinfo{volume}{58}},
  \bibinfo{pages}{5} (\bibinfo{year}{1936}).

\bibitem[{\citenamefont{Chomsky}(1956)}]{chomsky1956three}
\bibinfo{author}{\bibfnamefont{N.}~\bibnamefont{Chomsky}},
  \bibinfo{journal}{Information Theory, IRE Transactions on}
  \textbf{\bibinfo{volume}{2}}, \bibinfo{pages}{113} (\bibinfo{year}{1956}).

\bibitem[{\citenamefont{Lloyd}(1993)}]{lloyd1993quantum}
\bibinfo{author}{\bibfnamefont{S.}~\bibnamefont{Lloyd}},
  \bibinfo{journal}{Physical review letters} \textbf{\bibinfo{volume}{71}},
  \bibinfo{pages}{943} (\bibinfo{year}{1993}).

\bibitem[{\citenamefont{Cubitt et~al.}(2015)\citenamefont{Cubitt, Perez-Garcia,
  and Wolf}}]{cubitt2015undecidability}
\bibinfo{author}{\bibfnamefont{T.~S.} \bibnamefont{Cubitt}},
  \bibinfo{author}{\bibfnamefont{D.}~\bibnamefont{Perez-Garcia}},
  \bibnamefont{and} \bibinfo{author}{\bibfnamefont{M.~M.} \bibnamefont{Wolf}},
  \bibinfo{journal}{Nature} \textbf{\bibinfo{volume}{528}},
  \bibinfo{pages}{207} (\bibinfo{year}{2015}).

\bibitem[{\citenamefont{Solomonoff}(1964)}]{solomonoff1964formal}
\bibinfo{author}{\bibfnamefont{R.~J.} \bibnamefont{Solomonoff}},
  \bibinfo{journal}{Information and control} \textbf{\bibinfo{volume}{7}},
  \bibinfo{pages}{1} (\bibinfo{year}{1964}).

\bibitem[{\citenamefont{Levin}(1974)}]{levin1974laws}
\bibinfo{author}{\bibfnamefont{L.}~\bibnamefont{Levin}},
  \bibinfo{journal}{Problemy Peredachi Informatsii}
  \textbf{\bibinfo{volume}{10}}, \bibinfo{pages}{30} (\bibinfo{year}{1974}).

\bibitem[{\citenamefont{Delahaye and Zenil}(2012)}]{delahaye2012numerical}
\bibinfo{author}{\bibfnamefont{J.}~\bibnamefont{Delahaye}} \bibnamefont{and}
  \bibinfo{author}{\bibfnamefont{H.}~\bibnamefont{Zenil}},
  \bibinfo{journal}{Appl. Math. Comput.} \textbf{\bibinfo{volume}{219}},
  \bibinfo{pages}{63} (\bibinfo{year}{2012}).

\bibitem[{\citenamefont{Zenil et~al.}(2014)\citenamefont{Zenil, Soler-Toscano,
  Dingle, and Louis}}]{zenil2014correlation}
\bibinfo{author}{\bibfnamefont{H.}~\bibnamefont{Zenil}},
  \bibinfo{author}{\bibfnamefont{F.}~\bibnamefont{Soler-Toscano}},
  \bibinfo{author}{\bibfnamefont{K.}~\bibnamefont{Dingle}}, \bibnamefont{and}
  \bibinfo{author}{\bibfnamefont{A.~A.} \bibnamefont{Louis}},
  \bibinfo{journal}{Physica A: Statistical Mechanics and its Applications}
  \textbf{\bibinfo{volume}{404}}, \bibinfo{pages}{341} (\bibinfo{year}{2014}).

\bibitem[{\citenamefont{Soler-Toscano et~al.}(2014)\citenamefont{Soler-Toscano,
  Zenil, Delahaye, and Gauvrit}}]{soler2014calculating}
\bibinfo{author}{\bibfnamefont{F.}~\bibnamefont{Soler-Toscano}},
  \bibinfo{author}{\bibfnamefont{H.}~\bibnamefont{Zenil}},
  \bibinfo{author}{\bibfnamefont{J.-P.} \bibnamefont{Delahaye}},
  \bibnamefont{and} \bibinfo{author}{\bibfnamefont{N.}~\bibnamefont{Gauvrit}},
  \bibinfo{journal}{PloS one} \textbf{\bibinfo{volume}{9}},
  \bibinfo{pages}{e96223} (\bibinfo{year}{2014}).

\bibitem[{\citenamefont{Dingle et~al.}(2018)\citenamefont{Dingle, Camargo, and
  Louis}}]{dingle2018input}
\bibinfo{author}{\bibfnamefont{K.}~\bibnamefont{Dingle}},
  \bibinfo{author}{\bibfnamefont{C.~Q.} \bibnamefont{Camargo}},
  \bibnamefont{and} \bibinfo{author}{\bibfnamefont{A.~A.} \bibnamefont{Louis}},
  \bibinfo{journal}{Nature communications} \textbf{\bibinfo{volume}{9}},
  \bibinfo{pages}{761} (\bibinfo{year}{2018}).

\bibitem[{\citenamefont{Zenil et~al.}(2019)\citenamefont{Zenil, Badillo,
  Hern{\'a}ndez-Orozco, and Hern{\'a}ndez-Quiroz}}]{zenil2019coding}
\bibinfo{author}{\bibfnamefont{H.}~\bibnamefont{Zenil}},
  \bibinfo{author}{\bibfnamefont{L.}~\bibnamefont{Badillo}},
  \bibinfo{author}{\bibfnamefont{S.}~\bibnamefont{Hern{\'a}ndez-Orozco}},
  \bibnamefont{and}
  \bibinfo{author}{\bibfnamefont{F.}~\bibnamefont{Hern{\'a}ndez-Quiroz}},
  \bibinfo{journal}{International Journal of Parallel, Emergent and Distributed
  Systems} \textbf{\bibinfo{volume}{34}}, \bibinfo{pages}{161}
  (\bibinfo{year}{2019}).

\bibitem[{\citenamefont{P{\'e}rez et~al.}(2018)\citenamefont{P{\'e}rez, Louis,
  and Camargo}}]{perez2018deep}
\bibinfo{author}{\bibfnamefont{G.~V.} \bibnamefont{P{\'e}rez}},
  \bibinfo{author}{\bibfnamefont{A.~A.} \bibnamefont{Louis}}, \bibnamefont{and}
  \bibinfo{author}{\bibfnamefont{C.~Q.} \bibnamefont{Camargo}},
  \bibinfo{journal}{arXiv preprint arXiv:1805.08522}  (\bibinfo{year}{2018}).

\bibitem[{\citenamefont{Mingard et~al.}(2019)\citenamefont{Mingard, Skalse,
  Valle-Pérez, Martínez-Rubio, Mikulik, and Louis}}]{mingard2019neural}
\bibinfo{author}{\bibfnamefont{C.}~\bibnamefont{Mingard}},
  \bibinfo{author}{\bibfnamefont{J.}~\bibnamefont{Skalse}},
  \bibinfo{author}{\bibfnamefont{G.}~\bibnamefont{Valle-Pérez}},
  \bibinfo{author}{\bibfnamefont{D.}~\bibnamefont{Martínez-Rubio}},
  \bibinfo{author}{\bibfnamefont{V.}~\bibnamefont{Mikulik}}, \bibnamefont{and}
  \bibinfo{author}{\bibfnamefont{A.~A.} \bibnamefont{Louis}},
  \bibinfo{journal}{Arxiv preprint arXiv:1909.11522}  (\bibinfo{year}{2019}).

\bibitem[{\citenamefont{Valle-P{\'e}rez
  et~al.}(2018)\citenamefont{Valle-P{\'e}rez, Camargo, and
  Louis}}]{valle2018deep}
\bibinfo{author}{\bibfnamefont{G.}~\bibnamefont{Valle-P{\'e}rez}},
  \bibinfo{author}{\bibfnamefont{C.~Q.} \bibnamefont{Camargo}},
  \bibnamefont{and} \bibinfo{author}{\bibfnamefont{A.~A.} \bibnamefont{Louis}},
  \bibinfo{journal}{arXiv preprint arXiv:1805.08522}  (\bibinfo{year}{2018}).

\bibitem[{\citenamefont{Rosenblatt}(1958)}]{rosenblatt1958perceptron}
\bibinfo{author}{\bibfnamefont{F.}~\bibnamefont{Rosenblatt}},
  \bibinfo{journal}{Psychological review} \textbf{\bibinfo{volume}{65}},
  \bibinfo{pages}{386} (\bibinfo{year}{1958}).

\bibitem[{\citenamefont{LeCun et~al.}(2015)\citenamefont{LeCun, Bengio, and
  Hinton}}]{lecun2015deep}
\bibinfo{author}{\bibfnamefont{Y.}~\bibnamefont{LeCun}},
  \bibinfo{author}{\bibfnamefont{Y.}~\bibnamefont{Bengio}}, \bibnamefont{and}
  \bibinfo{author}{\bibfnamefont{G.}~\bibnamefont{Hinton}},
  \bibinfo{journal}{nature} \textbf{\bibinfo{volume}{521}},
  \bibinfo{pages}{436} (\bibinfo{year}{2015}).

\bibitem[{\citenamefont{G{\'a}cs}(1988)}]{gacs1988lecture}
\bibinfo{author}{\bibfnamefont{P.}~\bibnamefont{G{\'a}cs}},
  \emph{\bibinfo{title}{Lecture notes on descriptional complexity and
  randomness}} (\bibinfo{publisher}{Boston University, Graduate School of Arts
  and Sciences, Computer Science Department}, \bibinfo{year}{1988}).

\bibitem[{\citenamefont{Chaitin}(1974)}]{chaitin1974information}
\bibinfo{author}{\bibfnamefont{G.}~\bibnamefont{Chaitin}},
  \bibinfo{journal}{IEEE Transactions on Information Theory}
  \textbf{\bibinfo{volume}{20}}, \bibinfo{pages}{10} (\bibinfo{year}{1974}).

\bibitem[{\citenamefont{Calude et~al.}(2011)\citenamefont{Calude, Salomaa, and
  Roblot}}]{calude2011finite}
\bibinfo{author}{\bibfnamefont{C.~S.} \bibnamefont{Calude}},
  \bibinfo{author}{\bibfnamefont{K.}~\bibnamefont{Salomaa}}, \bibnamefont{and}
  \bibinfo{author}{\bibfnamefont{T.~K.} \bibnamefont{Roblot}},
  \bibinfo{journal}{Theoretical Computer Science}
  \textbf{\bibinfo{volume}{412}}, \bibinfo{pages}{5668} (\bibinfo{year}{2011}).

\bibitem[{\citenamefont{Gr{\"u}nwald and Roos}(2019)}]{grunwald2019minimum}
\bibinfo{author}{\bibfnamefont{P.}~\bibnamefont{Gr{\"u}nwald}}
  \bibnamefont{and} \bibinfo{author}{\bibfnamefont{T.}~\bibnamefont{Roos}},
  \bibinfo{journal}{arXiv preprint arXiv:1908.08484}  (\bibinfo{year}{2019}).

\end{thebibliography}

\end{document}


\title{{\LARGE Supplementary Information for: \\ \large Generic predictions of output probability based on complexities of inputs and outputs}}

\author{Kamaludin Dingle$^{1,2}$, Guillermo Valle P\'erez$^2$, Ard A. Louis$^2$}

\affiliation{$^1$Centre for Applied Mathematics and Bioinformatics,
Department of Mathematics and Natural Sciences,
Gulf University for Science and Technology, Kuwait,\\
$^2$Rudolf Peierls Centre for Theoretical Physics$\text{,}$ University of Oxford$\text{,}$ Parks Road,
Oxford, OX1 3PU, United Kingdom}

\date{\today}
\maketitle


\section{RNA sequence to secondary structure mapping}

RNA  is made of a linear sequence of 4  different kinds of nucleotides, so that there are $N_I=4^L$ possible sequences for any particular length $L$. A versatile molecule, it can  store information, as messenger RNA, or else  perform catalytic or structural functions.     For functional RNA, the three-dimensional  (3D) structure plays an important role in its function.  In spite of decades of research, it remains difficult to reliably predict the 3D structure from the sequence alone.  However, there are fast and accurate algorithms to calculate the so-called secondary structure (SS) that determines which base binds to which base.   Given a sequence, these methods typically minimize  the Turner model~\cite{mathews2004incorporating}  for the free-energy of a particular bonding pattern.  The main contributions in the Turner model are the hydrogen bonding and stacking interactions between the nucleotides, as well as some entropic factors to take into account motifs such as loops.  Fast algorithms based on dynamic programming allow for rapid calculations of these SS, and so this mapping from sequences to SS has been a popular model for many studies in biophysics.

  In this context, we view it as an input-output map, from $N_I$ input sequences to $N_O$ output SS structures.   This map has been extensively studied (see e.g.~\cite{schuster1994sequences, hofacker1994fast,fontana2002modelling,cowperthwaite2008ascent,aguirre2011topological,schaper2014arrival,wagner2005robustness,wagner2011origins,dingle2015structure}) and provided  profound insights into the biophysics of folding and evolution.

 Here we use the popular Vienna package~\cite{hofacker1994fast} to fold sequences to structures, with all parameters set to their default values (e.g.\ the temperature $T = 37^{\circ}C$). We folded all $N_I = 4^{15} \approx 10^9$ sequences of length $15$, into $346$ different structures which were the free-energy minimum structures for those sequences.    The number of sequences mapping to a structure is often called the \emph{neutral set size}.

 The structures can be abstracted in standard dot-bracket notation, where brackets denote bonds, and dots denote unbonded pairs.  For example, $...((....))....$. means that the first three bases are not bonded, the fourth and fifth are bonded, the sixth through ninth are unbonded, the tenth base is bonded to the fifth base, the eleventh base is bonded to the fourth base, and  the final four bases are unbounded.

 To estimate the complexity of an RNA SS, we first converted the dot-bracket representation of the structure into a binary string $x$, and then used the complexity estimators of section~\ref{sec:complexity}  to estimate its complexity.  To convert to binary strings, we replaced each dot with the bits 00, each left-bracket with the bits 10, and each right-bracket with 01. Thus an RNA SS of length $n$ becomes a bit-string of length $2n$. As an example, the following $n=15$ structure yields the displayed 30-bit string
\[
..(((...)))....\rightarrow 000010101000000001010100000000
\]

Because we are interested in exhaustive calculations, we are limited to rather small RNA sequence lengths.   This means that finite-size effects may play an important role.   In~\cite{dingle2018input},  we compared the simplicity bias bound ~(1) from the main text to longer sequences where only partial sampling can be achieved, and showed much clearer simplicity bias is evident in those systems.

\section{Finite state transducer}



Finite state transducers (FSTs) are a generalization of finite state machines that produce an output. They are defined by a finite set of states $\mathcal{S}$, finite input and output alphabets $\mathcal{I}$ and $\mathcal{O}$, and a transition function $T: \mathcal{S}\times\mathcal{I} \to \mathcal{S}\times\mathcal{O}$ defining, for each state, and input symbol, a next state, and output symbol. One also needs to define a distinguished state, $S_0\in\mathcal{S}$, which will be the initial state, before any input symbol has been read. Given an \emph{input sequence} of $L$ input symbols, the system visits different states, and simultaneously produces an \emph{output sequence} of $L$ output symbols.

FST form a popular toy system for computable maps. They can express any computable function that requires only a finite number of memory, and the number of states in the FSTs offers a good parameter to control the complexity of the map.  The class of machines we described above is also known as Mealy machines \cite{holcombe2004algebraic}. If one restricts the transition function to only depend on the current state, one obtains Moore machines~\cite{moore1956gedanken}. If one considers the input sequence to a Moore machine to be stochastic, it immediately follows that its state sequence follows a Markov chain, and its output sequence is a Markov information source. Therefore, FSTs can be used to model many stochastic systems in nature and engieering, which can be described by finite-state Markov dynamics.


FSTs lie in the lowest class in the Chomsky hierarchy. However, they appear to be biased towards simple outputs in a manner similar to Levin's coding theorem. In particular,  Zenil et al.~\cite{zenil2019coding} show evidence of this by correlating the probability of FSTs and UTMs producing particular outputs. More precisely, they sampled random FSTs with random inputs, and random UTMs with random inputs, and then compared the empirical frequencies with with individual output strings are obtained by both families, after many samples of machines and inputs.  For both types of machines, simple strings were much more likely to be produced than complex strings.

We use randomly generated FSTs with $5$ states. The FSTs are generated by uniformly sampling complete initially connected DFAs (where every state is reachable from the initial state, and the transition function is defined for every input) using the library FAdo~\cite{almeida2009fado}, which uses the algorithm developed by Almeida et al.~\cite{almeida2007enumeration}. Output symbols are then added to each transition independently and with uniform probability. In our experiments, the inputs are binary strings and the outputs are binary strings of length $L=30$. The outputs for the whole set of $2^L$ input strings are computed using the HFST library (\url{https://hfst.github.io/}). Not all FSTs show bias, but we have observed that all those that show bias show simplicity bias, and have the same behavior as that shown in Figure~\ref{fig:simpbias_FST} for low complexity - low probability outputs.

We can see why some simple outputs will occur with low probability by considering system specific details of the FST. For an FST, an output of length $n$ which is $n/2$ zeros followed by $n/2$ ones is clearly simple, but we find that it has a low probability. We can understand this intuitively as follows. Producing such a string requires the ``counting'' up to $n/2$ to know when to switch output, and counting requires a memory that grows with $n$, while FSTs have finite memory. We can also prove that, for instance, an FST that only produces such strings (for any $n$) is impossible. The set of possible strings that an FST can produce comprises a regular language, as constructed by using the output symbols at each transition as input symbols, giving us a non-deterministic finite automaton. Finally, using the pumping lemma \cite{rabin1959finite}, it is easy to see that this family of strings isn't a regular language.

\section{Perceptron}

The perceptron \cite{rosenblatt1958perceptron} is the simplest type of artificial neural network. It consists of a single linear layer, and a single output neuron with binary activation.  Because modern deep neural network architectures are typically made of many layers of perceptrons, this simple system is important to study~\cite{mingard2019neural}.   In this paper we use perceptrons with Boolean inputs and discretized weights. For inputs $x \in \{0,1\}^n$, the discretized perceptron uses the following parametrized class of functions:

\[
f_{w,b} (x) = \mathbf{1}(w \cdot x + b),
\]
where $w\in\{-a,a+\delta,\dots,a-\delta,a\}^n$ and $b\in\{-a,a+\delta,\dots,a-\delta,a\}$ are the weight vector and bias term, which take values in a discrete lattice with $D:=2a/\delta+1$ possible values per weight. We used $D=2^k$, so that each weight can be represented by $k$ bits, and $a=(2^k-1)/2$, so that $\delta=1$. Note that rescaling all the weights $w$ and the bias $b$ by the same fixed constant wouldn't change the family of functions.

To obtain the results in Figure~\ref{fig:simpbias_perceptron}, for which $n=7$, we represented the weights and bias with $k=3$ bits. We exhausitvely enumerated all $2^{3(7+1)}$ possible values of the weights and vectors, and we counted how many times we obtained each possible Boolean function on the Boolean hypercube $\{0,1\}^7$. The weight-bias pair was represented using $3\times(7+1)=24$ bits. A pair $(w,b)$ is an input to the parameter-function map of the perceptron. The complexity of inputs to this map can therefore be approximated by computing the Lempel-Ziv complexity of the 24-bit representation of the pair $(w,b)$.

\begin{figure*}[htp]
\centering
\subfigure[]{\label{fig:edge-d}\includegraphics[height=6cm,width=6cm]{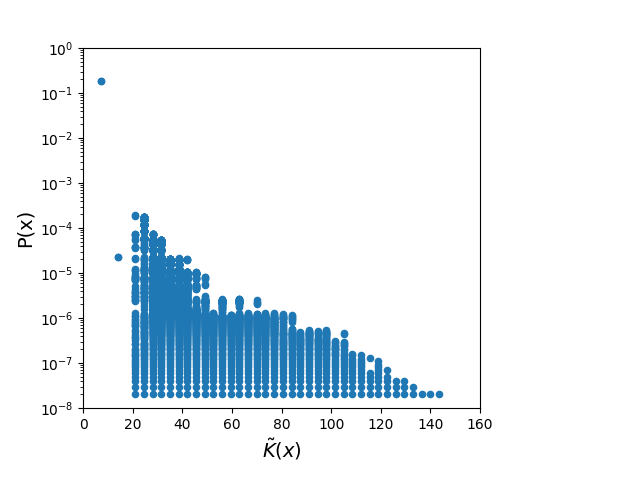}}
\subfigure[]{\label{fig:edge-d}\includegraphics[height=6cm,width=6cm]{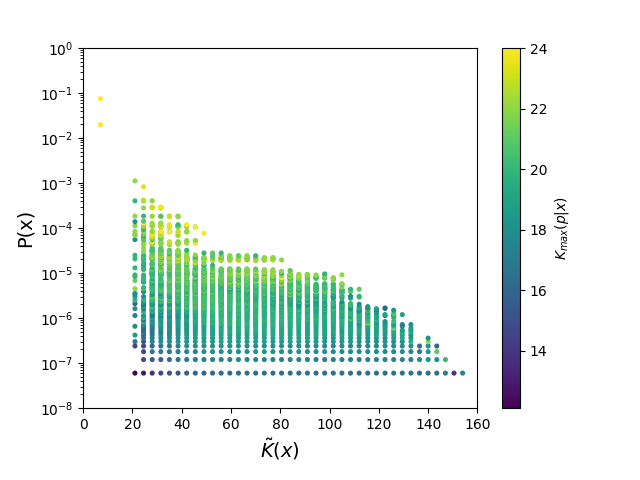}}
\caption{Probability versus complexity $\tilde{K}(x)$ (measured here as $C_{LZ}(x)$ from Eq.~(\ref{eq:CLZ})) shows simplicity bias in the perceptron for (a) full continuous weights and (b) with discretised weights.  Since weights and biases are real-valued in Fig.~(a) it is not straightforward to measure the complexity of the inputs, as it is for the discretised weights of Fig.~(b).   }
\label{fig:fulllperceptron}
\end{figure*}

In Figure~\ref{fig:fulllperceptron}, we compare the simplicity bias of a perceptron with real-valued weights and bias, sampled from a standard Normal distribution, to the simplicity bias of the perceptron with discretized weights. We observe that both display similar simplicity bias, although the profile of the upper bound changes slightly.

For the perceptron we can also understand some simple examples of low complexity, low probability outputs. For example, the function with all 0s except a 1, for the inputs (1,0,0,0,0,0,0) and  (0,1,0,0,0,0,0) has a similar complexity to the function which only has 1s at the inputs (1,0,0,0,0,0,0) and (0,1,1,1,1,1,1). However, the latter has much lower probability. One can understand this because if we take the dot product of a random weight vector $w$ with two different inputs $x_1$ and $x_2$, the results have correlation given $x_1 \cdot x_2 / (||x_1||||x_2||)$. Therefore we expect the input (0,1,1,1,1,1,1) to be correlated to more other inputs, than (0,1,0,0,0,0,0), so that the probability of it having a different value than the majority of inputs (as is the case for the second function) is expected to be significantly lower.

\section{Methods to estimate complexity $\tilde{K}(x)$}\label{sec:complexity}
\subsection{Lempel-Ziv compression}
There is a much more extensive discussion of different ways to estimate the Kolmogorov complexity in the supplementary information of~\cite{dingle2018input} and~\cite{perez2018deep}.   Here we use compression based measures, and as in these previous papers, we  these are based on the 1976 Lempel Ziv (LZ) algorithm~\cite{lempel1976complexity}, but with some small changes:
\begin{equation}
C_{LZ}(x) =\begin{cases}
     \log_2(n), &  \hspace*{-0.3cm}  \text{$x=0^n$ or $1^n$}\\
    \log_2(n) [N_w(x_1...x_n) + N_{w}(x_n...x_1)]/2, & \hspace*{-0.2cm} \text{otherwise}
  \end{cases}\label{eq:CLZ}
\end{equation}
Here  $N_{w}(x)$ is the number of code words found by the LZ algorithm.
The reason for distinguishing $0^n$ and $1^n$ is merely an artefact of $ N_{w}(x)$ which assigns complexity $K=1$ to the string 0 or 1, but complexity 2 to $0^n$ or $1^n$ for $n\geq2$, whereas the Kolmogorov complexity of such a trivial string actually scales as $\log_2(n)$, as one only needs to encode $n$.   In this way we ensure that our $C_{LZ}(x)$ measure not only gives the correct behaviour for complex strings in the $\lim_{n \rightarrow \infty}$, but also the correct behaviour for the simplest strings. In addition to the $\log_2(n)$ correction, taking the mean of the complexity of the forward and reversed strings makes the measure more fine-grained, since it allows more values for the complexity of a string.
 Note that $C_{LZ}(x)$ can also be used for strings of larger alphabet sizes than just 0/1 binary alphabets.

\subsection{Scaling complexities}

To directly test the input based measures we typically need fairly small systems, where the LZ based measure above may show some anomalies (see also the supplementary information of~\cite{dingle2018input} for a more detailed description).  Thus, for such small systems, or when comparing different types and sizes of objects (e.g.\ RNA SS and RNA sequences) a slightly different scaling may be more appropriate, which not only accounts for the fact that $C_{LZ}(x)>n$ for strings of length $n$, but also the lower complexity limit may not be $\sim0$, which it should be (see also the discussion in the supplementary information of~\cite{dingle2018input}).  Hence we use a different rescaling of the complexity measure
\begin{equation}
\tilde K(x) = \log_2(N_O)\cdot \frac{C_{LZ}(x) - \min(C_{LZ}(x))  }{\max(C_{LZ}(x)) - \min(C_{LZ}(x))}
\end{equation}
which will now range between $0\leq \tilde K(x) \leq  \log_2(N_O)=n$ if for example  $N_O=2^n$. 
For large objects, this different scaling will reduce to the simpler one, because $\max(C_{LZ}(x)) \gg \min(C_{LZ}(x))$. 

We note that there is nothing fundamental about using LZ to generate approximations to true Kolmogorov complexity.  Many other approximations could be used, and their merits may depend on the details of the problems involved.  For further discussion of other complexity measures, see for example the supplementary information of Refs.~\cite{dingle2018input,valle2018deep}.

\section{An alternative way to derive the cummulative bound}

Here we examine other ways of deriving what are effectively lower bounds on the probability, as expressed in the cumulative bound~(8).
First consider, as in~\cite{dingle2018input},  the function
\begin{equation}
q(x)=\frac{P_0(x)}{P(x)}  
\end{equation}
where 
$P_0(x) = 2^{-K(x|f,n)+\mathcal{O}(1)}$.  Here $q(x)$
 measures the ratio of the upper bound of equation~(\ref{eq:SB}) to the  probability $P(x)$ that an output $x$  is generated by random sampling of inputs.
Because we work with computable maps, $\sum_x P(x) = 1$, by definition. However, the bound $P_0(x)$ is not normalised, as it is an upper bound on the true probability.  One measure of its cumulative tightness is  to calculate the expected value of $q(x)$ summed over all inputs, which we call $\mathcal{E}_I$. This  can be written as a sum over all outputs, where every output is weighed as $P(x)$:
\begin{equation}
\mathcal{E}_I = \frac{1}{N_I} \sum_{i=1}^{N_I} q(x(p_i))=\sum_{j=1}^{N_O} P(x_j)q(x_j) =
\sum_{j=1}^{N_O} P_0(x_j) \label{eq:E}
\end{equation}
  By definition of an upper bound,  $q(x) \geq 1$ which means that $\mathcal{E}_I=\sum_{x \in O} 2^{-K(x|f,n) + \mathcal{O}(1)} \geq 1$. Interestingly,  because  $K(x|f,n)$ is a prefix code, $\sum_{x \in O} 2^{-K(x|f,n)} \leq 1$. Therefore $\mathcal{E}_I > 1$ due to the  $\mathcal{O}(1)$ terms.

In~\cite{dingle2018input} Markov's inequality was used to derive a lower bound upon uniform random sampling of inputs,
\begin{equation}\label{eq:lowerbnd}
\frac{P_0(x)}{\mathcal{E}_I r}\leq P(x) \leq P_0(x)
\end{equation}
which holds with a probability of at least $1-\frac{1}{r}$. The upper bound, given approximately by equation~(\ref{eq:SB}), always holds of course.
We measured $\mathcal{E}_I$ explicitly for the maps
in the main text compared to our approximate upper bound and
find that typically $\log_{10} \mathcal{E}_I \approx$ $1$ or $2$, which means that the bound is tight on a log scale at least.

Another related way to derive a cumulative bound such as that of Eq~(\ref{eq:sumP_delta}) follows a very simple argument. Recall that $\mathcal{D}(f)$ is defined as the set of all outputs $x_i$  that satisfy   $(\log_2(P_0(x_i)) - \log_2(P_(x_i))) \geq \Delta$. Recall also that the upper bound is defined as $P_0(x) = 2^{-K(x|f,n)+\mathcal{O}(1)}$. Then we can obtain the bound as follows.

\begin{align*}
\sum_{x \in \mathcal{D}(f)} P(x) &\leq \sum_{x \in \mathcal{D}(f)} P_0(x)2^{-\Delta} = \sum_{x \in \mathcal{D}(f)} 2^{-K(x|f,n)+\mathcal{O}(1)-\Delta} \\
&= 2^{-\Delta+\mathcal{O}(1)} \sum_{x \in \mathcal{D}(f)} 2^{-K(x|f,n)}\\
&\leq 2^{-\Delta+\mathcal{O}(1)} \sum_{x} 2^{-K(x|f,n)}\\
&\leq 2^{-\Delta+\mathcal{O}(1)},
\end{align*}
where the  last line follows from Kraft inequality~\cite{li2008introduction}, which applies because $K(x)$ comprise a prefix code.  If instead Eq~(\ref{eq:E}) were used for $\sum_x P_0(x) = \mathcal{E}_I$ in the derivation above, then we would obtain
\begin{equation}
\sum_{x \in \mathcal{D}(f)} P(x) \leq  \mathcal{E}_I  2^{-\Delta}
\end{equation}

  Although these arguments result in essentially the same bound  as the cumulative bound in the main text,  the  connection with the complexity of inputs is more opaque.  However, these derivations highlight other aspects of the bound, such as the role of the  $\mathcal{O}(1)$ term in the exponent. Therefore, the two derivations may give insight into the tightness of the looseness of the bound in different situations.
